\documentclass[twocolumn,showpacs,preprintnumbers,amsmath,amssymb]{revtex4}
\usepackage{graphicx}
\usepackage{dcolumn}
\usepackage{bm}
\usepackage[english]{babel}
\usepackage[utf8]{inputenc}
\usepackage{textcomp}

\begin{document}

\title{Coupling of optical far-fields into aperture-less plasmonic nanofibre tips}

\author{David Auw\"arter}
\author{Claus Zimmermann}
\author{Sebastian Slama}

\affiliation{Physikalisches Institut and Center for Collective Quantum Phenomena in LISA+, Universit\"at
T\"ubingen, Auf der Morgenstelle 14, D-72076 T\"ubingen, Germany}
\date{\today}

\begin{abstract}
This paper reports on the excitation of surface plasmons on gold-coated nanofibre tips by side-illumination with a laser beam and the coupling of the surface plasmons to the optical fiber. The measurements show a strong dependence of the coupling efficiency on the incidence angle with a maximum coupling efficiency on the order of few percent. Moreover, the fibre tip was used as scanning probe device for measuring the beam waist of a focussed laser beam. This work is motivated by the goal to use such plasmonic nanofibre tips in quantum optics experiments with cold atoms.
\end{abstract}

\pacs{}

\maketitle

\section{Introduction}
The idea to use plasmonic systems in quantum optics experiments has recently attracted much attention. Optical near-fields on plasmonic structures can be strongly enhanced and highly localized which makes them promising candidates for the generation of nanoscale potentials for cold atoms \cite{Chang09, Hau09, Stehle11}. Atoms which are trapped in such potentials can strongly couple to plasmonic excitations \cite{Lukin12} and interact with each other in a way that can be controlled by the trapping potential. The latter feature might provide a novel tool for the engineering of strong p-wave interactions, exotic quantum Hall states and topological gauge fields \cite{Diaz13}. In this context, we are investigating metallic nanotips with the perspective to trap single atoms close to the tip apex and couple their emission into surface plasmons with high Purcell enhancement \cite{Chang07}.\\
Metallic nanotips are often used in optical near field microscopy for obtaining sub-diffraction limited resolution and high sensitivity \cite{Novotny06}. While aperture tips are typically applied in a collection mode in which optical near-fields are coupled through a small sub-wavelength hole to an optical fibre, aperture-less metal tips are mostly used in a scattering mode in which an incoming laser beam is plasmonically enhanced and scattered at the tip apex. Recently, proposal have been put forward for using aperture-less tips also in collection mode \cite{Hecht05}. The idea is based on the coupling of the light modes of a tapered optical fibre with surface plasmons propagating on the metallic coating of the fibre, with a predicted conversion efficiency on the order of 10\% \cite{Ding07}. With a more elaborate setup the conversion efficiency could even be improved to almost 100\% \cite{Chen09}.\\ 
This paper deals with tapered dielectric glass fibre tips which are coated with a thin gold layer. We concentrate on the aspect of exciting surface plasmons on the tip by illumination with a laser beam. Moreover, we measure the coupling of the plasmons into guided modes of the fibre and find that a coupling efficiency on the order of few percent can be obtained. 

\section{Coupling of surface plasmons with free-space radiation}
\begin{figure*}[!t]
\begin{minipage}[btp]{0.4\textwidth}
 \begin{flushleft}
  a)
 \end{flushleft}
 \centerline{\includegraphics[height=5cm]{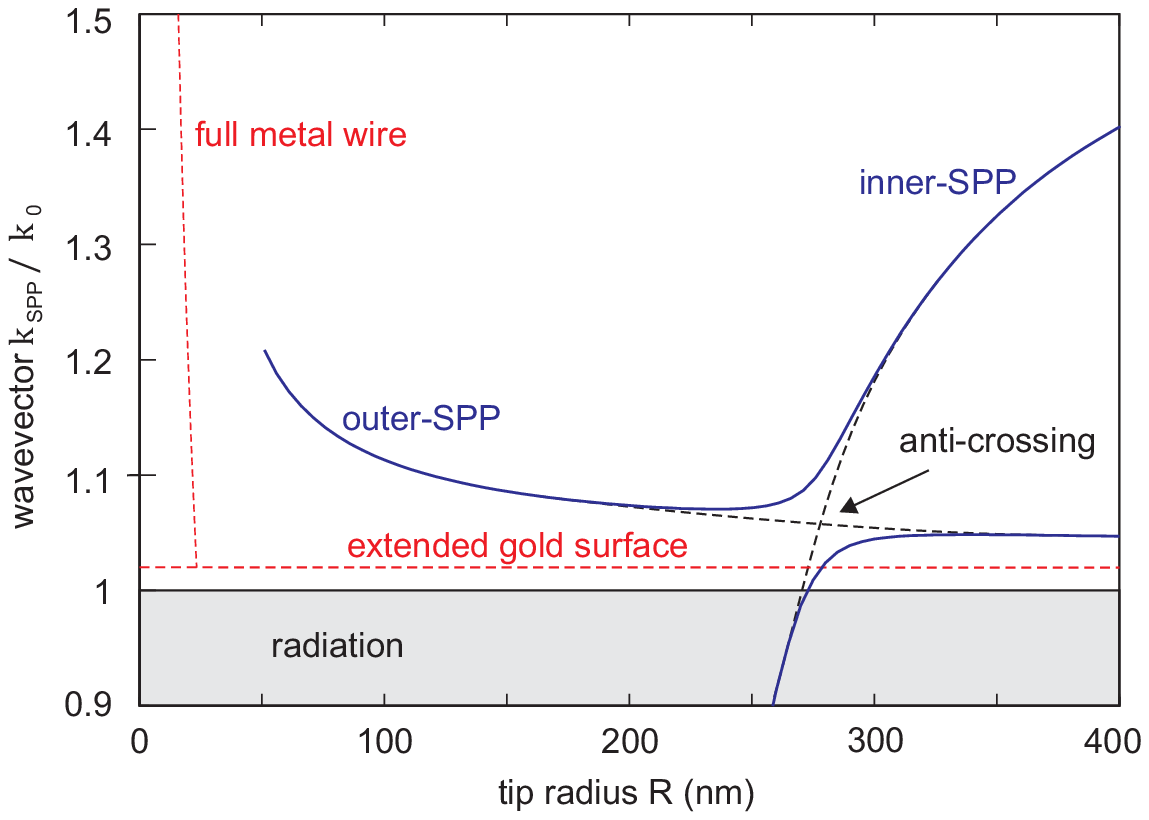}}

 \end{minipage}
 \begin{minipage}[btp]{0.4\textwidth}
 \begin{flushleft}
  b)
 \end{flushleft}
 \centerline{\includegraphics[height=5cm]{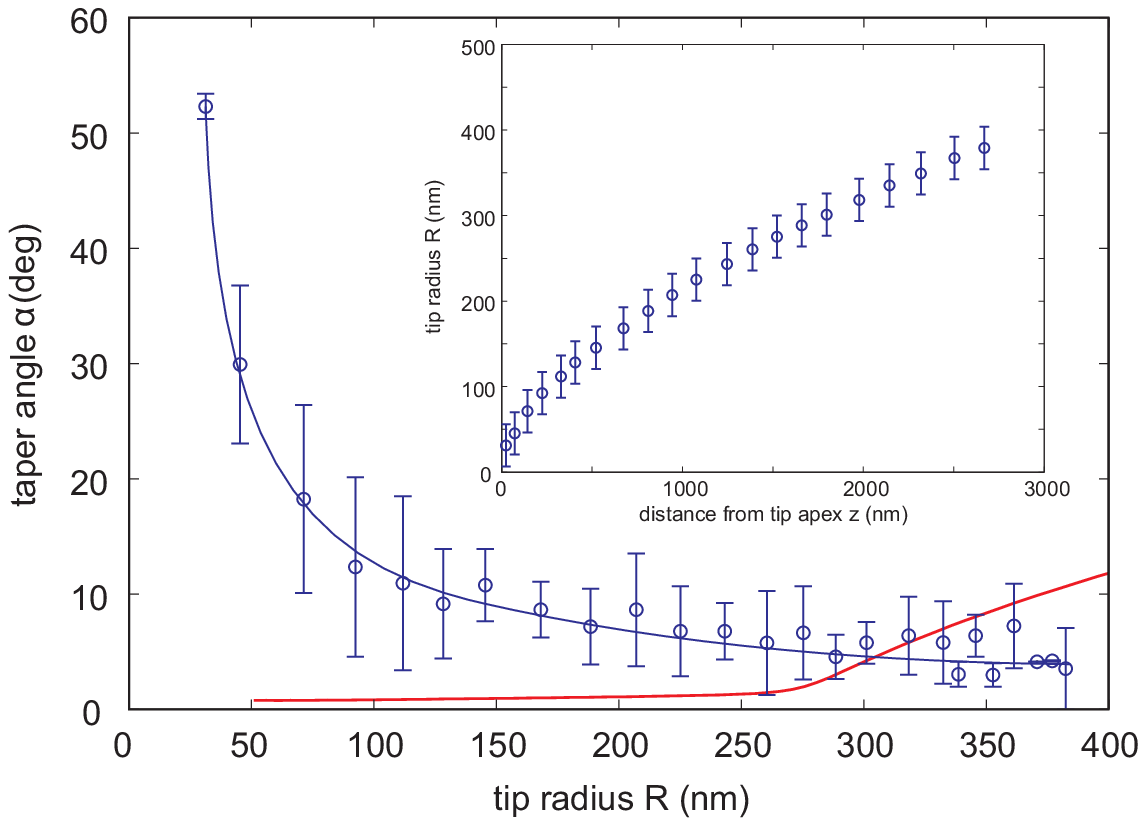}}
 \end{minipage}
 
\caption{a) Real part of the wavevector of surface plasmons (free-space wavelength $\lambda_0=780~\mathrm{nm}$) on the inner and outer surface of a $t=50~\mathrm{nm}$ thick metal film covering a tapered dielectric nanofibre with (uncoated) radius  $R-t$ and refractive index $n=1.5$. The coupling between inner-SPP and outer-SPP leads to an anticrossing. For tip radii $R<50~\mathrm{nm}$, the dielectric core is no longer present and the tip apex resembles a full metal wire. For comparison, we included also the wavevector of SPP on an extended thick gold layer. b) Surface plasmons can couple to free space radiation, if the real taper angle $\alpha$ (blue data points) is larger than the critical taper angle $\alpha_0(z)$ (red curve). The blue curve is for clarity only. The inset shows the experimentally determined tip radius as function of distances from the tip apex.}
\label{fig:theory}
\end{figure*}
Coupling of light modes requires in general an intersection of the dispersion relations of the participating modes. Here, the relevant dispersion relations are that of surface plasmons (SPP) propagating on the tip surface and that of free-space radiation with wavelength $\lambda_0$ and wavenumber $k_0=2\pi/\lambda_0$. The dispersion relation of surface plasmons propagating on a cylindrical  (full) metal wire with radius $R$ can be expressed analytically for small radii ($k_0R\ll1$) as \cite{Babadjanyan00, Stockman04, Stockman11}

\begin{eqnarray}\label{eq:spp_dispersion1}
&k_{\textrm{SPP}}&(\omega,k_0R\ll1)=\frac{1}{R}\times\\  & \times &\nonumber \left(-\frac{\epsilon_\textrm{m}(\omega)}{2\epsilon_\textrm{d}}\left(\ln\left(\sqrt{-\frac{4\epsilon_\textrm{m}(\omega)}{\epsilon_\textrm{d}}}\right)-\gamma\right)\right)^{-\frac{1}{2}}~.
\end{eqnarray}

In this equation the dielectric constant of the metal is $\epsilon_\textrm{m}=-24.1+i \cdot 1.72$ for gold at a wavelength of $\lambda_0=780~\mathrm{nm}$, the dielectric constant of the surrounding medium (air) is $\epsilon_\textrm{d}=1$, and $\gamma\approx0.57$ denotes the Euler-Mascheroni constant. For large radii $k_0R\gg1$ the dispersion relation approximates that of surface plasmons on planar surfaces \cite{Stockman04}

\begin{equation}\label{eq:spp_dispersion2}
k_{\textrm{SPP}}(\omega,k_0R\gg1)=\frac{\omega}{c}\sqrt{\frac{\epsilon_d\cdot\epsilon_m}{\epsilon_d+\epsilon_m}} ~.
\end{equation}

In the intermediate regime a gradual transistion connects both dispersion relations such that the wavenumber $k_{\textrm{SPP}}$ exceeds the free-space value $k_0$ for all radii. This prevents any coupling between surface plasmons and free space radiation. In the case of a dielectric fibre which is coated with a thin metal film, the situation is more complicated. SPP can exist on both the inner and the outer side of the metal film and are respectively called inner-SPP and outer-SPP. For a sufficiently thin metal layer the mutual coupling between inner-SPP and outer-SPP leads to an anticrossing \cite{Ding07, AlBader92}, see Fig \ref{fig:theory} a). Coupling to free-space radiation can in principle occur in a short range of tip radii for inner-SPP, i.e. in the lower branch of the anticrossing. However, the main reason for radiation coupling of SPP on plasmonic nanofibre tips is due to the taper of the tip \cite{Ding07}. SPP propagating on the tip surface can follow their dispersion relation adiabatically, only if the tip radius is changing not too fast, respectively if the local taper angle is not too large. A rough condition for adiabaticity of SPP propagation is given by \cite{Ding07}
 
\begin{equation}\label{eq:adiabaticity}
\alpha(z) < \alpha_0(z)=R(z) \mathrm{Re}\left[\frac{k_\mathrm{SPP}(z)}{k_0}-1 \right]/\lambda_0~,
\end{equation}

with $\alpha_0(z)$ the critical taper angle. In Fig. \ref{fig:theory} b) the critical taper angle $\alpha_0$ is shown in comparison with the real taper angle as determined from a scanning electron microscopy (SEM)-picture of the nanotip. The experimental taper angle is larger than the critical value at taper radii below $R\lesssim 300 ... 350~\mathrm{nm}$.  As shown in the inset of Fig. \ref{fig:theory} b) this corresponds to a distance range from the tip apex of $z\sim0...2~\mu\mathrm{m}$. At such distances coupling between SPP and radiation is possible due to the unadiabaticity of SPP propagation.

\section{Experimental setup}
\begin{figure*}[!t]
\begin{minipage}[btp]{0.4\textwidth}
 \begin{flushleft}
  a)
 \end{flushleft}
 \centerline{\includegraphics[height=5cm]{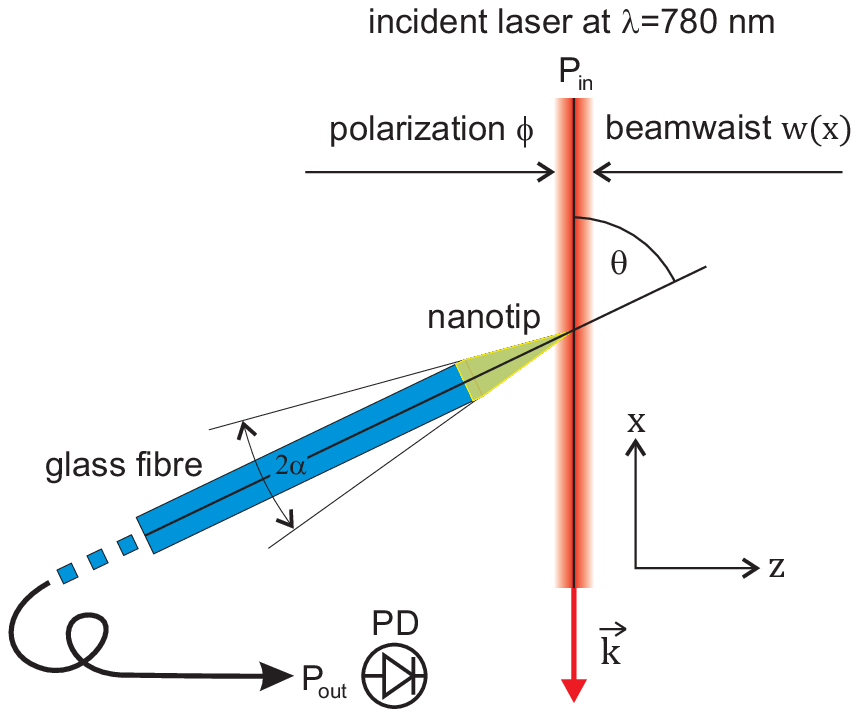}}

 \end{minipage}
 \begin{minipage}[btp]{0.4\textwidth}
 \begin{flushleft}
  b)
 \end{flushleft}
\centerline{\includegraphics[height=5cm]{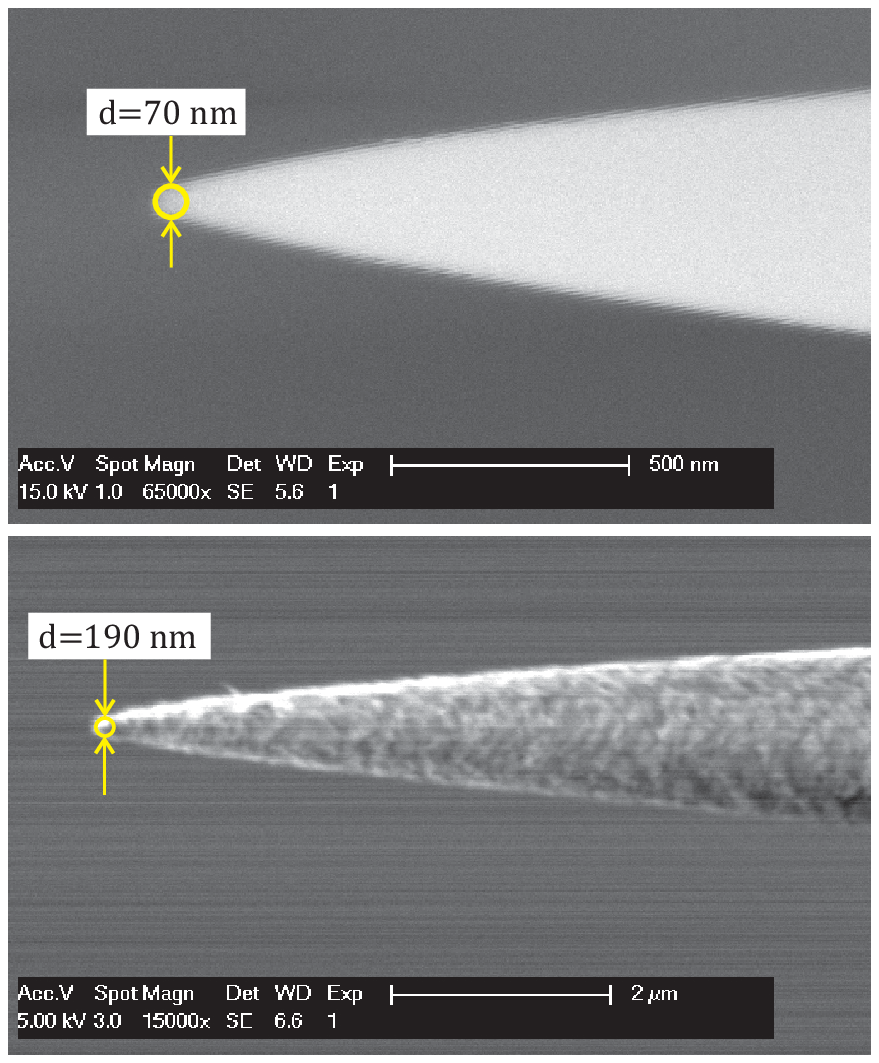}}
 \end{minipage}

 
\caption{a) Scheme of the experimental setup. The polarization $\phi$ is defined with respect to the surface of the fibre tip with $\theta=90^\circ$ - i.e. for s-polarization, the electric field of the laser is along the $z$-axis. b)Typical scanning electron microscopy (SEM)-pictures before and after the coating step.}
\label{fig:setup}
\end{figure*}

The experimental setup is shown in Fig. \ref{fig:setup}. Nanotips have been fabricated by pulling standard single mode glass fibres (Thorlabs S630-HP) with a commercial fibre pulling stage (Sutter Instruments P-2000). We obtain typical taper angles of $\alpha=\left(5\pm1\right)^\circ$ and typical curvature radii at the tip apex of $R_0=\left(35\pm5\right)~\mathrm{nm}$, see Fig. \ref{fig:setup} b). After the pulling step, the tip is coated with a thin adhesion layer of titanium (approx. $2.5~\mathrm{nm}$) and a gold layer with a typical thickness of $d=\left(50\pm10\right)~\mathrm{nm}$. The coating is done in a sputtering machine under permanent rotation of the fibre in order to reach a uniform gold layer thickness. The thickness is measured by comparison of SEM-pictures before and after the coating step. The metallized fibre tip is fixed to a combined rotation and nanopositioning stage (Attocube ECS3030) and illuminated from the side with a focussed laser beam, Fig. \ref{fig:setup} a). The wavelength of the laser $\lambda=780~\mathrm{nm}$ is chosen 
to correspond to the transition wavelength of the D2-line of Rb which is the most popular element in cold atom experiments. We measure the light power $P_\mathrm{out}$ which is emitted from the untapered fibre end with a photodiode (FDD-100) and amplify the signal with a transimpedance amplifier (FEMTO DLPCA-200). The absolute value of the emitted laser power is calibrated with a power meter (Thorlabs PM100A, S121C). It depends on the efficiency of several processes: (i) the excitation of surface plasmons on the tip, (ii) the conversion of surface plasmons into guided modes inside the fibre and (iii) the occurence of additional losses (e.g. reflection at the untapered end of the fibre). 

\section{Angular dependence}

\begin{figure*}[!t]
\begin{minipage}[btp]{0.4\textwidth}
 \begin{flushleft}
  a)
 \end{flushleft}
 \centerline{\includegraphics[height=5cm]{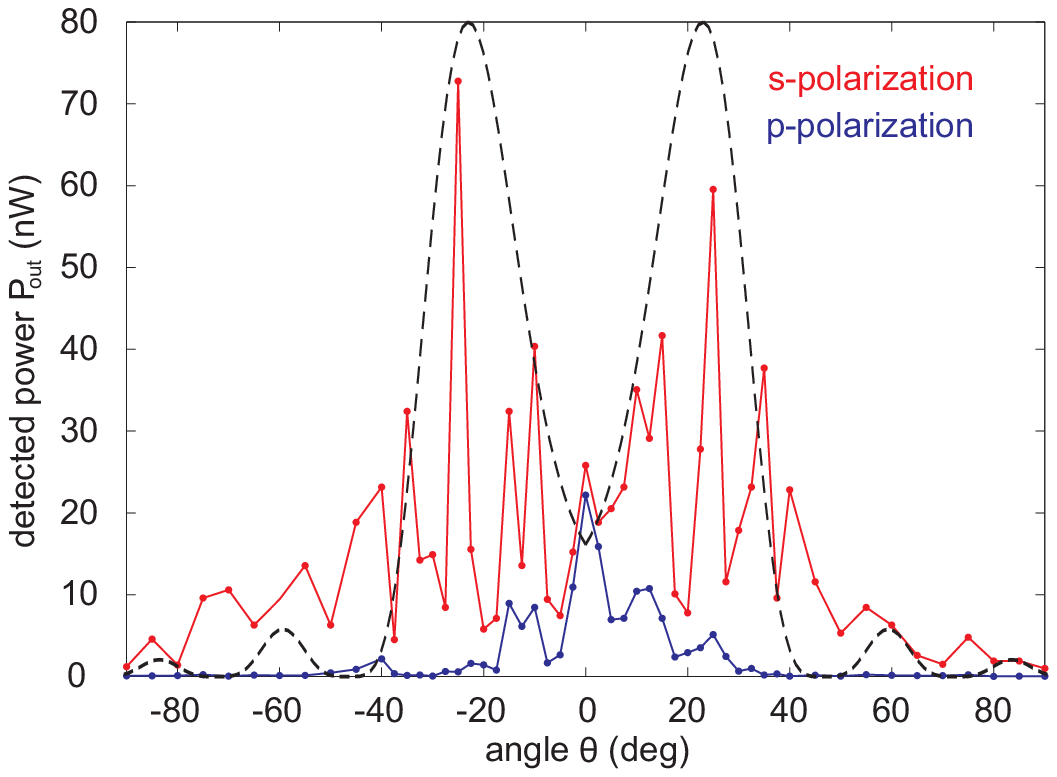}}
 \end{minipage}
 \begin{minipage}[btp]{0.4\textwidth}
 \begin{flushleft}
  b)
 \end{flushleft}
 \centerline{\includegraphics[height=5cm]{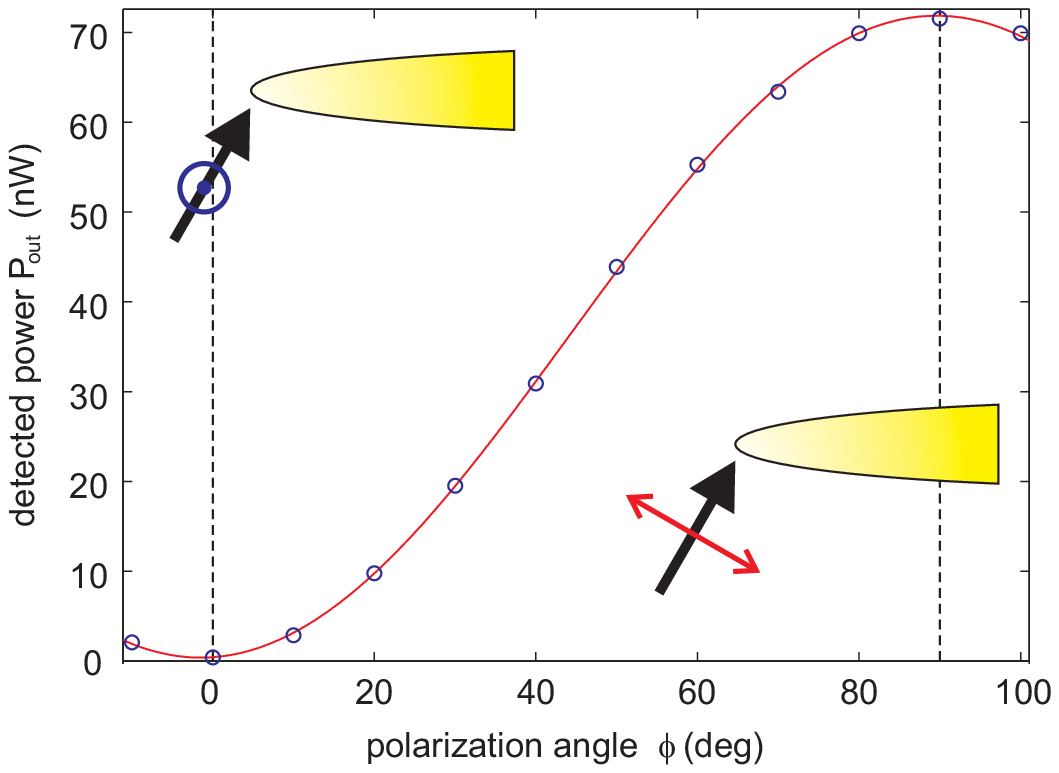}}
 \end{minipage}
 
\caption{a) Detected light power as a function of the angle $\theta$ between the incident laser and the fibre axis for both polarizations. The incident laser power is $P_\textrm{in}=1~\mathrm{mW}$. The dashed line describes the angular dependence as interference of plasmon waves at the position $z_c$ where outer-SPP are transferred into inner-SPP, with $z_c=1.9~\mu\mathrm{m}$ and a taper angle of $\alpha=5^\circ$. The height of the theoretical curve is adjusted to the figure scale. b) Detected light power as a function of the polarization $\phi$ of the incident laser with $\theta=25^\circ$.}
\label{fig:measurement}
\end{figure*}

The angle $\theta$ between the fibre tip and the laser beam axis is varied with the rotation stage. After each step of rotation the position of the tip is readjusted to the laser beam focus by maximizing the detected power with the translation stages. The measured light power which is emitted at the untapered fibre end is plotted in Fig. \ref{fig:measurement} a). Little power is coupled into the fibre if the laser beam is aligned perpendicularly to the tip axis ($\theta=90^\circ$). Suprisingly, very sharp maxima are observed at certain angles in the range between $\theta\approx10^\circ$ and $\theta\approx 40^\circ$. The general shape of this angle dependence can be understood in a simple model as interference between different plasmon waves that are excited along the (conical) nanotip. Depending on the incidence angle $\theta$ and the taper angle $\alpha$ they are launched with a position dependent relative phase $\phi(z)=k_0z\frac{\cos(\alpha+\theta)}{\cos(\alpha)}$. From there they propagate towards the position $z_c$ at which outer-SPP are transferred into inner-SPP. The phase at the transfer position  $z_c$ of a plasmon launched at $z$ is thus $\phi(z_c)=\frac{k_0}{\cos(\alpha)}\left[z_c-z\left(1-\cos(\theta+\alpha)\right)\right]$. For simplicity, we describe in this formula plasmon propagation with the free-space wavevector $k_0$. The sum over the electric fields of all participating waves $\Sigma_{z=0}^{z_c}\cos(\phi(z_c))$ leads to an angle dependent intensity $I(\theta)$ which is plotted in Fig. \ref{fig:measurement} a) as dashed line. The theoretical curve fits well to the envelope of the measured maxima, though it does not resolve the distinct resonances. We attribute the occurrence of these resonances to more complicated effects due to the exact shape of the nanotip.\\

As a cross-check for the importance of the direction of the eletric field oscillation, we changed the polarization of the laser beam from s to p. The corresponding measured laserpower at $\theta=25^\circ$ is reduced by a factor of one hundred, see Fig. \ref{fig:measurement} a). Gradually changing the polarization from s to p results in an almost perfect sinusoidal behavior, see Fig. \ref{fig:measurement} b). The fact that the curves in Fig. \ref{fig:measurement} a) are symmetric with respect to the $\theta=0^\circ$ axis is a signature for the good radial symmetry of the tip and its uniform coating with gold. For unsymmetrically coated tips (by switching off tip rotation during the sputtering process) we observed very unsymmetric excitation curves (not shown here). We would also like to note that the metal layer thickness on the nanotip is a crucial parameter for the signal height. For layer thickness deviations of $50\%$ from the ideal value, we observed signal heights  that were reduced by approximately one order of magnitude. 

\section{Coupling efficiency}

\begin{figure*}[!t]
\begin{minipage}[btp]{0.4\textwidth}
 \begin{flushleft}
  a)
 \end{flushleft}
 \centerline{\includegraphics[height=5cm]{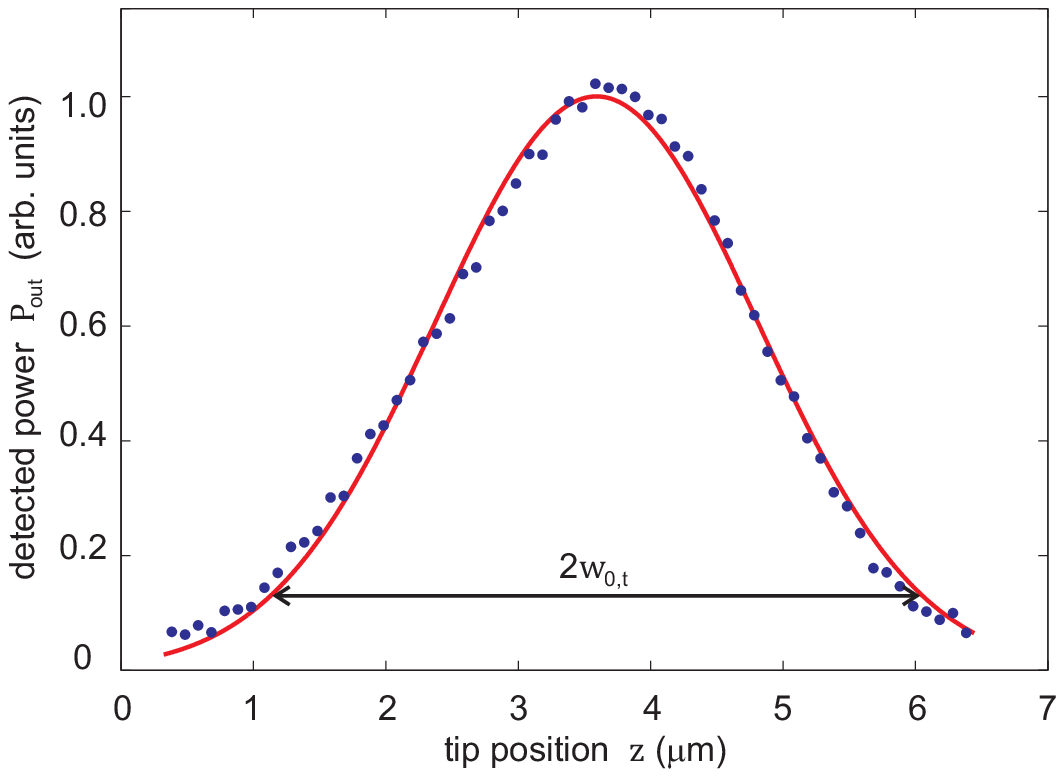}}
 \end{minipage}
 \begin{minipage}[btp]{0.4\textwidth}
 \begin{flushleft}
  b)
 \end{flushleft}
 \centerline{\includegraphics[height=5cm]{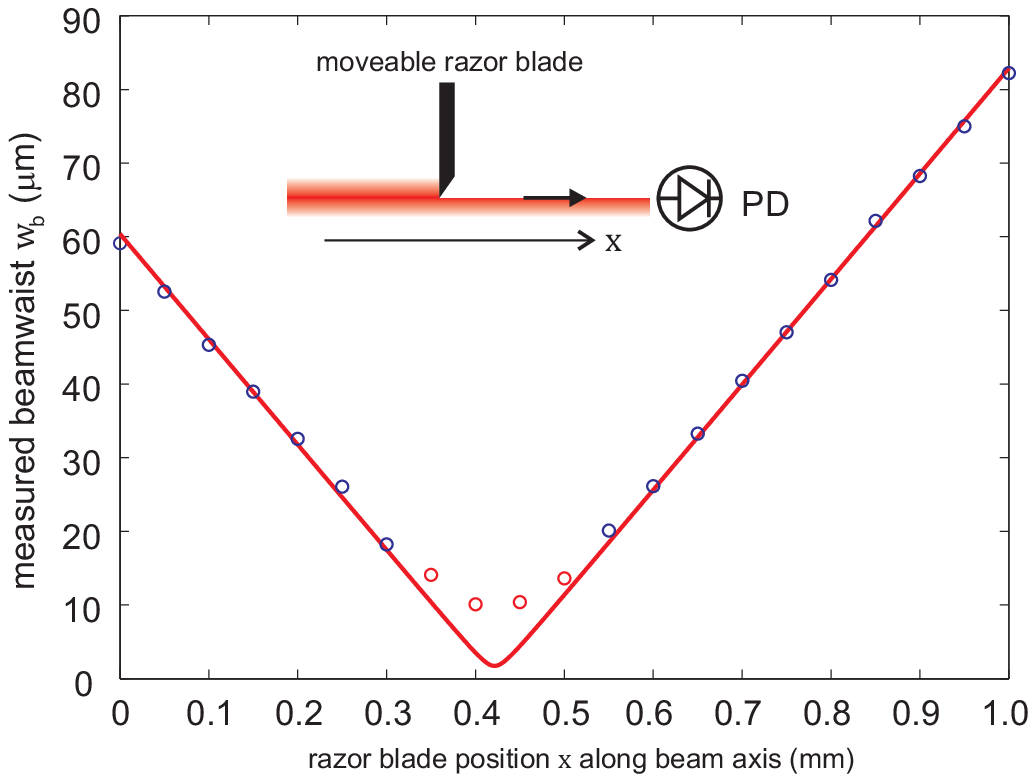}}
 \end{minipage}
 
\caption{a) The tip is used in a scanning mode for measuring the beamwaist of a focussed laser beam. (b) Independent measurement of the beam waist with a razor blade.}
\label{fig:beamwaist}
\end{figure*}

The coupling efficiency $\eta$ is defined as the ratio of the detected light power $P_\mathrm{out}$ to the power $P_\mathrm{tip}$ illuminating the zone on the nanotip, where plasmons can be excited.
\begin{equation}\label{eq:coupling}
\eta=\frac{P_\mathrm{out}}{P_\mathrm{tip}}~,
\end{equation}
with $P_\mathrm{tip}=I_0\cdot A$, the maximum intensity of the irradiated laser beam $I_0=2P_\mathrm{in}/\pi w_0^2$ and the area of the sensitive zone on the nanotip $A$.
The length of this zone $\Delta z$ is determined by measuring the beamwaist of a tightly focussed laser beam by displacing the nanotip perpendicularly to the beam axis with the nanopositioning stage. The detected laser power is shown in Fig. \ref{fig:beamwaist} a). The data points are fitted with a Gaussian curve $G(z)\sim\exp\left(-2z^2/w_\mathrm{0,t}^2\right)$ with waist $w_\mathrm{0,t}=\left(2.44\pm 0.05\right)~\mu\mathrm{m}$. The measured waist is given by the convolution of the Gaussian intensity distribution $I(z)\sim\exp\left(-2z^2/w_0^2\right)$ of a $\textrm{TEM}_{00}$ laser beam (beam waist $w_0$) with the excitation profile on the nanotip $f(z)$. For simplicity, we assume a Gaussian excitation profile $f(z)\sim\exp\left(-2z^2/\Delta z^2\right)$, for which the coupling length can be expressed analytically as 
\begin{equation}\label{eq:deltaz}
\Delta z=\sqrt{w_\mathrm{0,t}^2-w_0^2}~.
\end{equation}
The real beam waist $w_0$ is measured independently with a razor blade \cite{Eichler04}, see Fig. \ref{fig:beamwaist} b). As this method is not suitable for measuring small beamwaists due to diffraction at the razor blade, the beamwaist $w_0$ in the focal plane is obtained by fitting the measured beamwaists far away from the focus with Gaussian beam propagation in free space $w_\mathrm{b}(x)=w_0\sqrt{1+\left(x/x_0\right)^2}$, where the Rayleigh length is given by $x_0=\pi w_0^2/\lambda$. Neglecting the diffraction limited data points (red circles), the fit results in a beamwaist of $w_0=\left(1.74\pm 0.05\right)~\mu\mathrm{m}$, corresponding to a coupling length $\Delta z=\left(1.71\pm 0.07\right)~\mu\mathrm{m}$. Please note, that this value agrees very well with the coupling length as determined from Fig. \ref{fig:theory} b). The area of the sensitive zone on the tip is estimated as $A\sim\Delta z\cdot w_\textrm{tip}=\left(1.2\pm0.1\right)\times10^{-12}~\mathrm{m}^2$, where the width of the tip $w_\textrm{tip}=\left(0.7\pm0.05\right)~\mu\mathrm{m}$ is evaluated from a SEM image at a distance $\Delta z$ from the apex. Using the experimental parameters of the measurements shown in Fig. \ref{fig:measurement} ($P_\mathrm{in}=1~\mathrm{mW}$, $w_0=\left(23.2\pm0.4\right)~\mu\mathrm{m}$) the laser power on the sensitive zone is $P_\mathrm{tip}=\left(1.4\pm0.2\right)~\mu\mathrm{W}$. This value has to be compared with the maximum detected power of $P_\mathrm{max}=73~\mathrm{nW}$, corresponding to a coupling efficiency of 
$\eta=\left(5\pm1\right)\%$. This coupling efficiency involves both the excitation probability of plasmons on the surface and the conversion efficiency of surface plasmons to guided modes in the fibre. Please note, that the sensitive area $A$ is an estimate based on the assumption of a Gaussian excitation profile along the nanotip and a constant excitation profile across it. Thus, the obtained coupling efficiency rather indicates the order of magnitude than represents a precision measurement. Moreover, the coupling length was determined from a different tip specimen than the measurements shown in Fig. \ref{fig:measurement}. The same analysis with the tip specimen used for the measurements shown in Fig. \ref{fig:measurement} resulted in a beam waist (measured with the tip) of $w_\mathrm{0,t}=\left(23.4\pm 0.35\right)~\mu\mathrm{m}$ compared with a beamwaist (measured with the razor blade) of $w_0=\left(23.2\pm0.4\right)~\mu\mathrm{m}$). This corresponds to a coupling efficiency of $\eta=\left(2.6_{-2}^{+97.4}\right)\%$. The large uncertainty towards high coupling efficiencies stems from the fact that $w_\mathrm{0,t}<w_0$ cannot be excluded within measured precision. However, both results are consistent with a coupling efficiency on the order of few percent.

\section{Conclusion}
Concluding, we have shown that surface plasmons can be excited on a metal nanotip in a simple way by illumination with a laser beam from the side. This setup was proposed in \cite{Chang09} for the generation of nanoscale dipole traps for single atoms close to the tip apex. We have shown that the complex interaction of SPP on the inner and on the outer side of the gold layer lead to a adiabaticity criterium that allows excitations of SPP with far field radiation in a range of roughly 2 microns from the tip apex. On the other side, outer-SPP are transferred to inner-SPP at a tip radius of approximately $260~\mathrm{nm}$ by a broad anticrossing. Furthermore, the coupling of outer-SPP to inner-SPP has a complicated angular dependence on the irradiated laser beam which can be partly attributed to interference of propagating SPP. The detection of light at the end of the fibre allowed us also to estimate the coupling efficiency between the plasmon mode and guided modes within the fibre. We obtain a coupling efficiency on the order of few percent exceeding typical values of collection efficiency of standard NSOM setups \cite{Hecht00}. This high value is also promising for Purcell-enhanced detection of the spontaneous emission of atoms which are positioned close to the fibre tip.

\section{Acknowledgement}
S. Slama is indebted to the Baden-W\"urttemberg Stiftung for the financial support of this research project by the Eliteprogramm for Postdocs.

\section{Competing financial interests}
Competing financial interests do not exist.

\end{document}